\pdfoutput=1
\documentclass[12pt,a4paper]{article}

\usepackage{ifthen} 
\newboolean{pdflatex}
\setboolean{pdflatex}{true} 

\newboolean{articletitles}
\setboolean{articletitles}{true} 

\newboolean{uprightparticles}
\setboolean{uprightparticles}{false} 

\def\paperauthors{LHCb collaboration} 
\def\paperasciititle{Precision measurement of the Xib- baryon lifetime} 
\def\papertitle{Precision measurement of the \newline $\Xibm$ baryon lifetime} 
\def\paperkeywords{{High Energy Physics}, {LHCb}} 
\def\papercopyright{\the\year\ CERN for the benefit of the LHCb collaboration} 
\def\paperlicence{CC BY 4.0 licence}
\def\paperlicenceurl{https://creativecommons.org/licenses/by/4.0/}

\usepackage[top=1in, bottom=1.25in, left=1in, right=1in]{geometry}

\usepackage{microtype}
\usepackage{lineno}  
\usepackage{xspace} 
\usepackage{caption} 

\usepackage{graphicx}  
\usepackage{color}
\usepackage{colortbl}
\graphicspath{{./figs/}} 

\usepackage{amsmath} 
\usepackage{amssymb}
\usepackage{amsfonts}
\usepackage{upgreek} 

\newcommand*\patchAmsMathEnvironmentForLineno[1]{%
\expandafter\let\csname old#1\expandafter\endcsname\csname #1\endcsname
\expandafter\let\csname oldend#1\expandafter\endcsname\csname
end#1\endcsname
 \renewenvironment{#1}%
   {\linenomath\csname old#1\endcsname}%
   {\csname oldend#1\endcsname\endlinenomath}%
}
\newcommand*\patchBothAmsMathEnvironmentsForLineno[1]{%
  \patchAmsMathEnvironmentForLineno{#1}%
  \patchAmsMathEnvironmentForLineno{#1*}%
}
\AtBeginDocument{%
\patchBothAmsMathEnvironmentsForLineno{equation}%
\patchBothAmsMathEnvironmentsForLineno{align}%
\patchBothAmsMathEnvironmentsForLineno{flalign}%
\patchBothAmsMathEnvironmentsForLineno{alignat}%
\patchBothAmsMathEnvironmentsForLineno{gather}%
\patchBothAmsMathEnvironmentsForLineno{multline}%
\patchBothAmsMathEnvironmentsForLineno{eqnarray}%
}

\usepackage{hyperxmp}

\usepackage[pdftex,
            pdfauthor={\paperauthors},
            pdftitle={\paperasciititle},
            pdfkeywords={\paperkeywords},
            pdfcopyright={Copyright (C) \papercopyright},
            pdflicenseurl={\paperlicenceurl}]{hyperref}

\usepackage[colorinlistoftodos,textsize=scriptsize]{todonotes}

\usepackage[bottom,flushmargin,hang,multiple]{footmisc}

\usepackage[all]{hypcap} 

\usepackage{xspace} 
\usepackage{upgreek}

\def\lhcb   {\mbox{LHCb}\xspace}

\def\MagUp {\mbox{\em Mag\kern -0.05em Up}\xspace}

\ifthenelse{\boolean{uprightparticles}}%
{

 \def\Pmu         {\ensuremath{\upmu}\xspace}                 
 \def\Pnu         {\ensuremath{\upnu}\xspace}                 
                  
 \def\Ppi         {\ensuremath{\uppi}\xspace}

 \def\Ppsi        {\ensuremath{\uppsi}\xspace}

 \def\PDelta      {\ensuremath{\Delta}\xspace}                 
 \def\PXi         {\ensuremath{\Xi}\xspace}                 
 \def\PLambda     {\ensuremath{\Lambda}\xspace}                 
 \def\PSigma      {\ensuremath{\Sigma}\xspace}                 
 \def\POmega      {\ensuremath{\Omega}\xspace}                 
 \def\PUpsilon    {\ensuremath{\Upsilon}\xspace}
 \let\oldPi\Pi
 \def\PPi         {\ensuremath{\oldPi}\xspace}

 \def\PB      {\ensuremath{\mathrm{B}}\xspace}                 
                  
 \def\PD      {\ensuremath{\mathrm{D}}\xspace}

 \def\PJ      {\ensuremath{\mathrm{J}}\xspace}                 
 \def\PK      {\ensuremath{\mathrm{K}}\xspace}

 \def\Pb      {\ensuremath{\mathrm{b}}\xspace}                 
 \def\Pc      {\ensuremath{\mathrm{c}}\xspace}

 \def\Pi      {\ensuremath{\mathrm{i}}\xspace}

 \def\Ps      {\ensuremath{\mathrm{s}}\xspace}

 \def\thebaroffset{0.0em}
}
{

 \def\Pmu         {\ensuremath{\mu}\xspace}                 
 \def\Pnu         {\ensuremath{\nu}\xspace}                 
                  
 \def\Ppi         {\ensuremath{\pi}\xspace}

 \def\Ppsi        {\ensuremath{\psi}\xspace}                 
                  
 \mathchardef\PDelta="7101
 \mathchardef\PXi="7104
 \mathchardef\PLambda="7103
 \mathchardef\PSigma="7106
 \mathchardef\POmega="710A
 \mathchardef\PUpsilon="7107
 \mathchardef\PPi="7105
                  
 \def\PB      {\ensuremath{B}\xspace}                 
                  
 \def\PD      {\ensuremath{D}\xspace}

 \def\PJ      {\ensuremath{J}\xspace}                 
 \def\PK      {\ensuremath{K}\xspace}

 \def\Pb      {\ensuremath{b}\xspace}                 
 \def\Pc      {\ensuremath{c}\xspace}

 \def\Pi      {\ensuremath{i}\xspace}

 \def\Ps      {\ensuremath{s}\xspace}

 \def\thebaroffset{0.18em}
}
\newcommand{\offsetoverline}[2][\thebaroffset]{\kern #1\overline{\kern -#1 #2}}%

\makeatletter
\ifcase \@ptsize \relax
  \newcommand{\miniscule}{\@setfontsize\miniscule{4}{5}}
\or
  \newcommand{\miniscule}{\@setfontsize\miniscule{5}{6}}
\or
  \newcommand{\miniscule}{\@setfontsize\miniscule{5}{6}}
\fi
\makeatother

\DeclareRobustCommand{\optbar}[1]{\shortstack{{\miniscule (\rule[.5ex]{1.25em}{.18mm})}
  \\ [-.7ex] $#1$}}

\def\mun        {{\ensuremath{\Pmu^-}}\xspace} 

\def\neub       {{\ensuremath{\overline{\Pnu}}}\xspace}

\def\neumb      {{\ensuremath{\neub_\mu}}\xspace}

\def\squark    {{\ensuremath{\Ps}}\xspace}

\def\cquark    {{\ensuremath{\Pc}}\xspace}

\def\bquark    {{\ensuremath{\Pb}}\xspace}
\def\bquarkbar {{\ensuremath{\overline \bquark}}\xspace}
\def\bbbar     {{\ensuremath{\bquark\bquarkbar}}\xspace}

\def\pion   {{\ensuremath{\Ppi}}\xspace}

\def\pip    {{\ensuremath{\pion^+}}\xspace}
\def\pim    {{\ensuremath{\pion^-}}\xspace}

\def\kaon    {{\ensuremath{\PK}}\xspace}

\def\KorKbar {\kern \thebaroffset\optbar{\kern -\thebaroffset \PK}{}\xspace}

\def\Kp      {{\ensuremath{\kaon^+}}\xspace}
\def\Km      {{\ensuremath{\kaon^-}}\xspace}

\def\D       {{\ensuremath{\PD}}\xspace}

\def\DorDbar {\kern \thebaroffset\optbar{\kern -\thebaroffset \PD}\xspace}
\def\Dz      {{\ensuremath{\D^0}}\xspace}

\def\Dp      {{\ensuremath{\D^+}}\xspace}
\def\Dm      {{\ensuremath{\D^-}}\xspace}

\def\DpDm    {\ensuremath{\Dp {\kern -0.16em \Dm}}\xspace}

\def\Dstarp  {{\ensuremath{\D^{*+}}}\xspace}

\def\B       {{\ensuremath{\PB}}\xspace}

\def\BorBbar {\kern \thebaroffset\optbar{\kern -\thebaroffset \PB}\xspace}

\def\Bd      {{\ensuremath{\B^0}}\xspace}

\def\BdorBdbar {\kern \thebaroffset\optbar{\kern -\thebaroffset \Bd}\xspace}

\def\Bs      {{\ensuremath{\B^0_\squark}}\xspace}

\def\BsorBsbar {\kern \thebaroffset\optbar{\kern -\thebaroffset \Bs}\xspace}

\def\jpsi     {{\ensuremath{{\PJ\mskip -3mu/\mskip -2mu\Ppsi}}}\xspace}

\def\Y#1S{\ensuremath{\PUpsilon{(#1S)}}\xspace}

\def\Lz          {{\ensuremath{\PLambda}}\xspace}

\def\LorLbar     {\kern \thebaroffset\optbar{\kern -\thebaroffset \PLambda}\xspace}

\def\Xires       {{\ensuremath{\PXi}}\xspace}

\def\Xim         {{\ensuremath{\Xires^-}}\xspace}

\def\Omegares    {{\ensuremath{\POmega}}\xspace}

\def\Lc          {{\ensuremath{\Lz^+_\cquark}}\xspace}

\def\Xicz        {{\ensuremath{\Xires^0_\cquark}}\xspace}

\def\Lb           {{\ensuremath{\Lz^0_\bquark}}\xspace}

\def\Xibm         {{\ensuremath{\Xires^-_\bquark}}\xspace}

\def\Omegab       {{\ensuremath{\Omegares^-_\bquark}}\xspace}

\def\BF         {{\ensuremath{\mathcal{B}}}\xspace}
\def\BR         {\BF}

\def\to                 {\ensuremath{\rightarrow}\xspace}

\def\eff   {{\ensuremath{\varepsilon}}\xspace}

\def\AT#1     {\ensuremath{A_{\mathrm{T}}^{#1}}\xspace}           

\def\C#1      {\ensuremath{\mathcal{C}_{#1}}\xspace}                       
\def\Cp#1     {\ensuremath{\mathcal{C}_{#1}^{'}}\xspace}                    
\def\Ceff#1   {\ensuremath{\mathcal{C}_{#1}^{\mathrm{(eff)}}}\xspace}        
\def\Cpeff#1  {\ensuremath{\mathcal{C}_{#1}^{'\mathrm{(eff)}}}\xspace}       
\def\Ope#1    {\ensuremath{\mathcal{O}_{#1}}\xspace}                       
\def\Opep#1   {\ensuremath{\mathcal{O}_{#1}^{'}}\xspace}                    

\newcommand{\nospaceunit}[1]{\ensuremath{\text{#1}}}       
\newcommand{\aunit}[1]{\ensuremath{\text{\,#1}}}       

\newcommand{\tev}{\aunit{Te\kern -0.1em V}\xspace}
\newcommand{\gev}{\aunit{Ge\kern -0.1em V}\xspace}
\newcommand{\mev}{\aunit{Me\kern -0.1em V}\xspace}
\newcommand{\kev}{\aunit{ke\kern -0.1em V}\xspace}
\newcommand{\ev}{\aunit{e\kern -0.1em V}\xspace}
 
\newcommand{\mevc}{\ensuremath{\aunit{Me\kern -0.1em V\!/}c}\xspace}
\newcommand{\gevc}{\ensuremath{\aunit{Ge\kern -0.1em V\!/}c}\xspace}
\newcommand{\mevcc}{\ensuremath{\aunit{Me\kern -0.1em V\!/}c^2}\xspace}
\newcommand{\gevcc}{\ensuremath{\aunit{Ge\kern -0.1em V\!/}c^2}\xspace}

\def\mum  {\ensuremath{\,\upmu\nospaceunit{m}}\xspace}

\def\fb   {\ensuremath{\aunit{fb}}\xspace}
\def\invfb   {\ensuremath{\fb^{-1}}\xspace}

\def\ps   {\ensuremath{\aunit{ps}}\xspace}

\newcommand{\chisq}{\ensuremath{\chi^2}\xspace}

\newcommand{\chisqip}{\ensuremath{\chi^2_{\text{IP}}}\xspace}

\def\gsim{{~\raise.15em\hbox{$>$}\kern-.85em
          \lower.35em\hbox{$\sim$}~}\xspace}
\def\lsim{{~\raise.15em\hbox{$<$}\kern-.85em
          \lower.35em\hbox{$\sim$}~}\xspace}

\def\sPlot{\mbox{\em sPlot}\xspace}

\def\pt         {\ensuremath{p_{\mathrm{T}}}\xspace}

\def\ptot       {\ensuremath{p}\xspace}

\def\evtgen     {\mbox{\textsc{EvtGen}}\xspace}

\def\geant      {\mbox{\textsc{Geant4}}\xspace}

\def\photos     {\mbox{\textsc{Photos}}\xspace}

\def\pythia     {\mbox{\textsc{Pythia}}\xspace}

\def\tell1  {TELL1\xspace}
\def\ukl1   {UKL1\xspace}

\newcommand{\lhcborcid}[1]{\href{https://orcid.org/#1}{\hspace*{0.1em}\raisebox{-0.45ex}{\includegraphics[width=1em]{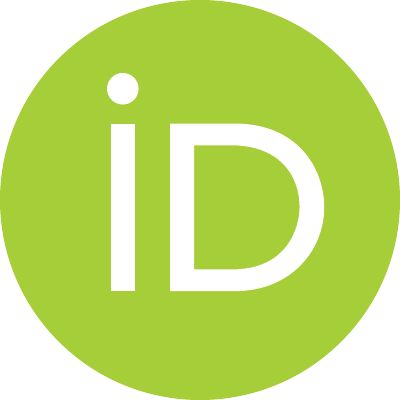}}}}

\usepackage{cite} 
\usepackage{mciteplus}

\usepackage{longtable} 

\begin{document}

\renewcommand{\thefootnote}{\fnsymbol{footnote}}
\setcounter{footnote}{1}


\begin{titlepage}
\pagenumbering{roman}

\vspace*{-1.5cm}
\centerline{\large EUROPEAN ORGANIZATION FOR NUCLEAR RESEARCH (CERN)}
\vspace*{1.5cm}
\noindent
\begin{tabular*}{\linewidth}{lc@{\extracolsep{\fill}}r@{\extracolsep{0pt}}}
\ifthenelse{\boolean{pdflatex}}
{\vspace*{-1.5cm}\mbox{\!\!\!\includegraphics[width=.14\textwidth]{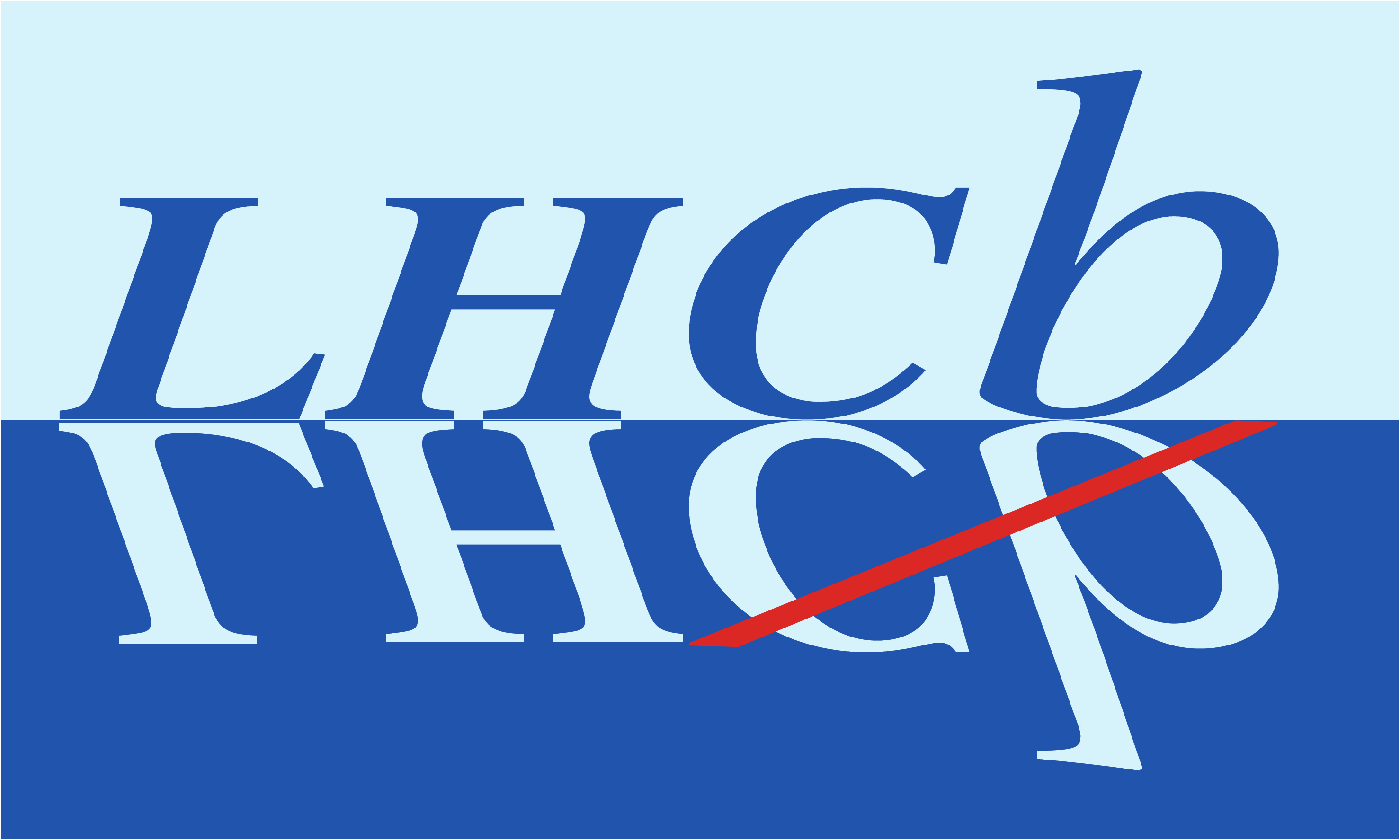}} & &}%
{\vspace*{-1.2cm}\mbox{\!\!\!\includegraphics[width=.12\textwidth]{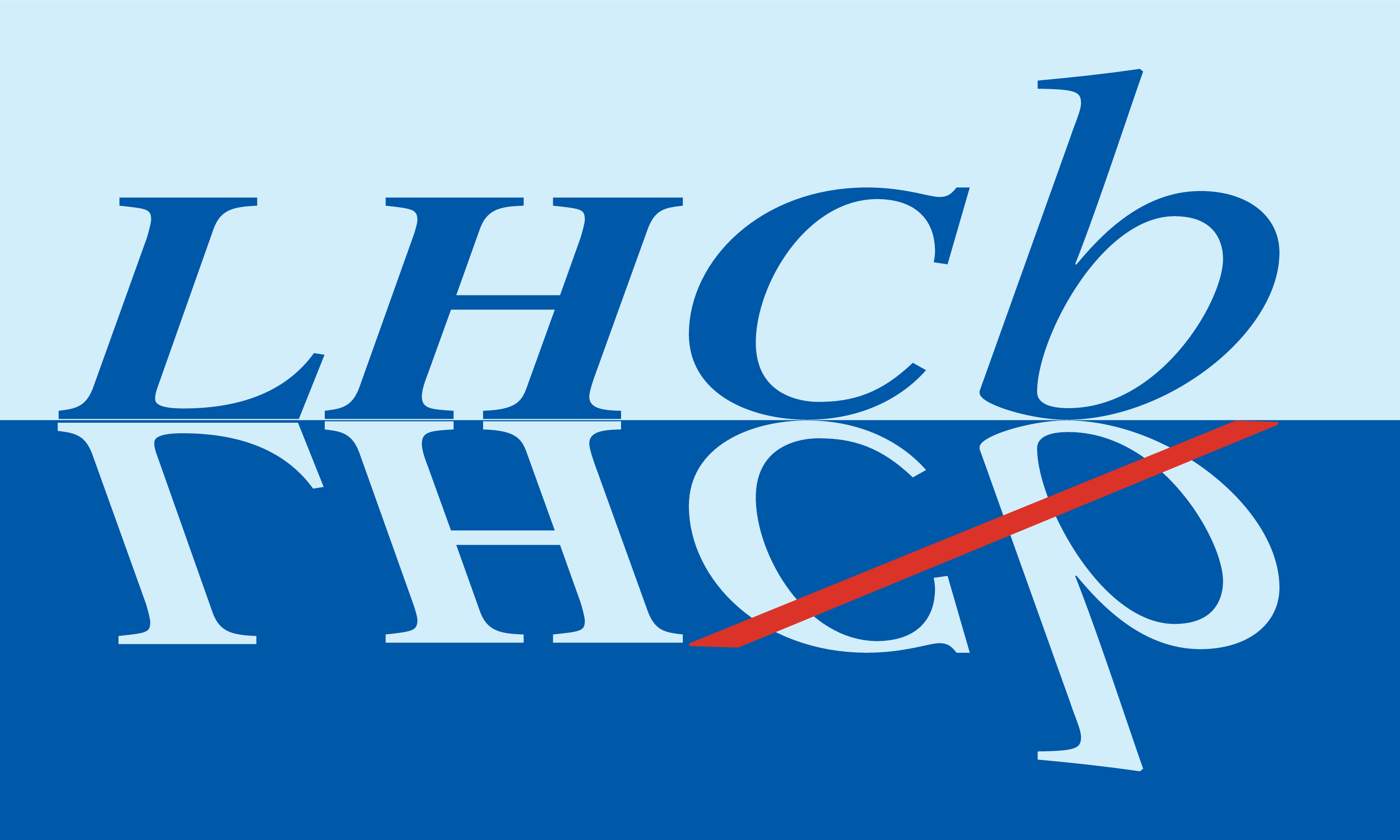}} & &}%
\\
 & & CERN-EP-2024-139 \\  
 & & LHCb-PAPER-2024-010 \\  
 & & June 17, 2024 \\ 
 & & \\
\end{tabular*}

\vspace*{4.0cm}

{\normalfont\bfseries\boldmath\huge
\begin{center}
  \papertitle 
\end{center}
}

\vspace*{2.0cm}

\begin{center}
\paperauthors\footnote{Authors are listed at the end of this paper.}
\end{center}

\vspace{\fill}

\begin{abstract}
  \noindent
A sample of $pp$ collision data, corresponding to an integrated luminosity of 5.5$\invfb$ and collected by the LHCb experiment during LHC Run 2, is used to measure the ratio of the lifetime of the $\Xibm$ baryon to that of the $\Lb$ baryon, $r_{\tau}\equiv\tau_{\Xibm}/\tau_{\Lb}$. The value ${r_{\tau}=1.076\pm0.013\pm0.006}$ is obtained, where the first uncertainty is statistical and the second systematic. 
This value is averaged with the corresponding value from Run 1 to obtain 
${r_{\tau}^{\rm Run\,1,2} = 1.078\pm0.012\pm0.007}$. Multiplying by the world-average value of the $\Lb$ lifetime yields ${\tau_{\Xibm}^{\rm Run~1,2} = 1.578\pm0.018\pm0.010\pm0.011~{\rm ps}}$, where the uncertainties are statistical, systematic, and due to the limited knowledge of the $\Lb$ lifetime. This measurement improves the precision of the current world average of the $\Xibm$ lifetime by about a factor of two, and is in good agreement with the most recent theoretical predictions.
\end{abstract}

\vspace*{2.0cm}

\begin{center}
  Published in
  Phys.~Rev.~D 
\end{center}

\vspace{\fill}

{\footnotesize 
\centerline{\copyright~\papercopyright. \href{\paperlicenceurl}{\paperlicence}.}}
\vspace*{2mm}

\end{titlepage}


\newpage
\setcounter{page}{2}
\mbox{~}


\renewcommand{\thefootnote}{\arabic{footnote}}
\setcounter{footnote}{0}

\cleardoublepage


\pagestyle{plain} 
\setcounter{page}{1}
\pagenumbering{arabic}


\section{Introduction}
\label{sec:Introduction}
The heavy quark expansion
(HQE)~\cite{Lenz:2014jha} is a theoretical framework that predicts the inclusive decay rates of beauty hadrons through an expansion in powers of the strong coupling constant, $\alpha_s$, and $\Lambda_{\rm QCD}/m_b$. Here, $\Lambda_{\rm QCD}$ is the energy scale at which the strong-interaction coupling becomes large, ${\mathcal{O}}(100\mev)$, and $m_b$ is the $b$-quark mass. When combined with precision measurements of heavy-quark decays, the HQE framework can be used to calculate $b$-hadron parameters required for the determination of Cabibbo-Kobayashi-Maskawa~\cite{Cabibbo:1963yz,Kobayashi:1973fv} matrix elements, which in turn provide constraints on physics beyond the Standard Model.

A stringent test of the HQE framework is to confront its predictions of lifetimes, {\it i.e.,} the inverse of the corresponding decay widths, with precision measurements. At leading order in the calculation using the HQE framework, the decay width of all $b$ hadrons is equal to that of the $b$ quark, and therefore, all $b$ hadrons have the same lifetime. The higher order corrections account for nonperturbative effects, the interactions of the $b$ quark with the spectator quarks, and the effects of Pauli interference~\cite{Neubert_Bdecays}, leading to differences in the total decay width of the $b$ hadrons. Therefore lifetime measurements provide a direct quantitative test of the HQE higher-order corrections. Improved theoretical calculations of lifetimes from the HQE have recently been published~\cite{Gratrex}, and they are generally in good agreement with the available measurements. 
The lifetime ratios of the $\Xibm$ and $\Omegab$ baryons with respect to that of the $\Lb$ baryon, 
$\tau_{\Xibm}/\tau_{\Lb}$ and $\tau_{\Omegab}/\tau_{\Lb}$, have theoretical uncertainties of $1.9\%$ and $4.2\%$, which are less than the experimental uncertainties of $2.5\%$ and $11\%$~\cite{PDG2022}, respectively.  

The $\Xibm$ baryon was first observed in 2007 by the CDF~\cite{CDF:2007cgg} and D0~\cite{D0:2007gjs} collaborations in proton-antiproton ($p\bar{p}$) collisions at $\sqrt{s}=1.96\tev$. The CDF collaboration subsequently measured the lifetime of the $\Xibm$ baryon using ${\Xibm\to\jpsi\Xim}$ decays~\cite{CDF:2014mon}. Soon after, measurements of the $\Xibm$ baryon lifetime were reported by the LHCb collaboration using $\Xibm\to\Xicz\pim$~\cite{LHCb-PAPER-2014-048} and $\Xibm\to\jpsi\Xim$~\cite{LHCb-PAPER-2014-010} decays, using proton-proton ($pp$) collision data taken between 2011 and 2012 (Run~1) at center-of-mass energies of $\sqrt{s}=7$ and 8\tev.  At LHCb, the $\Xibm\to\Xicz\pim$ decay mode provides about a factor of 2.5 better statistical precision than the ${\Xibm\to\jpsi\Xim}$ mode due to the higher total efficiency for detecting such decays.

In this paper, a new measurement of the $\Xibm$ baryon lifetime is reported using a $pp$ collision data sample collected between 2016 and 2018 (Run\,2) by the LHCb experiment at $\sqrt{s}=13\tev$, corresponding to an integrated luminosity of 5.5\invfb. The integrated luminosity and $\bbbar$ production cross section are both about a factor of two larger compared to the previous analysis~\cite{LHCb-PAPER-2014-048}.
The $\Xibm$ baryons are detected through their decays to the ${\Xicz\pim}$ final state, with ${\Xicz\to p\Km\Km\pip}$. The lifetime is normalized to that of the 
$\Lb$ baryon, which is selected through the decay $\Lb\to\Lc\pim$ with $\Lc\to p\Km\pip$. The inclusion of charge-conjugate processes is implied throughout this paper.

The quantity that is measured experimentally is the ratio of efficiency-corrected signal yields as a function of decay time, $t$,
\begin{align}
    R(t) \equiv\frac{N[\Xibm\to\Xicz\pim](t)}{N[\Lb\to\Lc\pim](t)} \cdot \frac{\eff[\Lb\to\Lc\pim](t)}{\eff[\Xibm\to\Xicz\pim](t)}=R_0\exp{(\lambda t)},
\label{eq:RT}
\end{align}
\noindent where $R_0$ is an overall normalization factor,  and $N$ and $\eff$ represent the signal yields and efficiencies of the decay mode indicated in the brackets, respectively. The parameter $\lambda$ is related to the lifetimes of the $\Lb$ and $\Xibm$ baryons through
\begin{align}
    \lambda\equiv\frac{1}{\tau_{\Lb}}-\frac{1}{\tau_{\Xibm}}.
\end{align}
\noindent The ratio of lifetimes is then
\begin{align}
    r_{\tau}\equiv\frac{\tau_{\Xibm}}{\tau_{\Lb}}=\frac{1}{1-\lambda\tau_{\Lb}}.
\label{eq:rtau}
\end{align}
\noindent Because $\lambda$ is small, about $0.05\ps^{-1}$, and $\tau_{\Lb}=1.464\pm0.010\ps$ is precisely measured~\cite{PDG2022}, the uncertainty in the $\Lb$ baryon lifetime is a very small contribution to the overall uncertainty in $r_{\tau}$.

\section{Detector and simulation}
The \lhcb detector~\cite{LHCb-DP-2008-001,LHCb-DP-2014-002} is a single-arm forward
spectrometer covering the \mbox{pseudorapidity} range $2<\eta <5$, designed for the study of particles containing \bquark or \cquark
quarks. The detector used for this analysis includes a high-precision tracking system
consisting of a silicon-strip vertex detector surrounding the $pp$
interaction region~\cite{LHCb-DP-2014-001}, a large-area silicon-strip detector located
upstream of a dipole magnet with a bending power of about
$4{\mathrm{\,T\,m}}$, and three stations of silicon-strip detectors and straw
drift tubes~\cite{LHCb-DP-2017-001} placed downstream of the magnet.
The tracking system provides a measurement of the momentum, \ptot, of charged particles with
a relative uncertainty that varies from 0.5\% at low momentum to 1.0\% at 200\gevc. The polarity of the LHCb magnet is alternated regularly throughout each period of data taking. The minimum distance of a track to a primary $pp$ collision vertex (PV), the impact parameter (IP), 
is measured with a resolution of approximately $\sigma_{\rm IP}=(15+29/\pt)\mum$,
where \pt is the component of the momentum transverse to the beam, in\,\gevc.
Different types of charged hadrons are distinguished using information
from two ring-imaging Cherenkov detectors~\cite{LHCb-DP-2012-003}. 
Photons, electrons and hadrons are identified by a calorimeter system consisting of
scintillating-pad and preshower detectors, an electromagnetic
and a hadronic calorimeter. Muons are identified by a
system composed of alternating layers of iron and multiwire
proportional chambers~\cite{LHCb-DP-2012-002}.
The online event selection is performed by a trigger~\cite{LHCb-DP-2019-001},  which consists of a hardware stage (L0), based on information from the calorimeter and muon systems, followed by a software stage, which applies a full event reconstruction. The software stage employs a multivariate algorithm~\cite{BBDT,LHCb-PROC-2015-018} to identify secondary vertices consistent with the decay of a \bquark hadron.

Simulation is required to model the effects of the detector acceptance and the imposed selection requirements. In the simulation, $pp$ collisions are generated using \pythia~\cite{Sjostrand:2007gs,*Sjostrand:2006za} 
with a specific \lhcb configuration~\cite{LHCb-PROC-2010-056}.
Decays of unstable particles are described by \evtgen~\cite{Lange:2001uf}, in which final-state radiation is generated using \photos~\cite{davidson2015photos}. The interaction of the generated particles with the detector, and its response, are implemented using the \geant toolkit~\cite{Allison:2006ve, *Agostinelli:2002hh} as described in Ref.~\cite{LHCb-PROC-2011-006}. The underlying $pp$ interaction is reused multiple times, with an independently generated signal decay each time~\cite{LHCb-DP-2018-004}. Unless otherwise indicated, the lifetimes of the beauty and charm baryons in the simulation use the world-average values~\cite{PDG2022}.
Simulated $\Xibm$ and $\Lb$ decays are generated for each of the three data-taking years --- 2016, 2017 and 2018 --- with conditions that correspond to the running conditions for that year. These conditions are primarily related to changes in the hardware and software trigger thresholds, implemented to adapt to changes in the instantaneous luminosity.

\section{Candidate selection}
\label{sec:selection}

Signal $\Xibm$ ($\Lb$) candidates are formed by combining in a kinematic fit a 
$\Xicz\to p\Km\Km\pip$ ($\Lc\to p\Km\pip$) candidate decay with a $\pim$ candidate. Hereafter, the notation $H_b$ and $H_c$ is used to refer to the $b$ and $c$ baryons, when the discussion applies to both the $\Lb$ and $\Xibm$ modes. 
All final-state particles from the $H_b$ baryons are required to have trajectories that are significantly detached from all PVs in the event, characterized by the quantity $\chisqip$. 
Its value is obtained from the difference in the vertex-fit $\chisq$ of a given PV reconstructed with and without the particle in question, and is approximately equal to $({\rm IP}/\sigma_{\rm IP})^2$. Thus, for particles originating from the $H_b$ or $H_c$ decay vertices, the $\chisqip$ values tend to be large, whereas for particles originating from the PV (such as the reconstructed $H_b$ baryon), they ought to be small.  The difference between the $\Xicz$ ($\Lc$) candidate mass and the known value~\cite{PDG2022} is required to be less than 15\mevcc (25\mevcc), which is about three times the mass resolution, and the $H_c$ decay time is required to satisfy $-0.5 < t_{\Xicz}< 1.245\ps$ ($-0.5 < t_{\Lc}< 2.5\ps$). The upper limits on the decay times are about ten times the known lifetimes~\cite{PDG2022}

Mass vetoes, in combination with additional particle identification (PID) requirements, are used to suppress crossfeeds from misidentified $D^{+}_{(s)}\to\Kp\Km\pip$, \mbox{$D^{*+}\to\Dz(\Kp\Km)\pip$}, and $\Dp\to\Km\pip\pip$ decays faking $\Lc\to p\Km\pip$ decays, as well as $\phi\to\Kp\Km$ decays in which a $\Kp$ meson is misidentified as a proton. These vetoes reduce the background in the $\Lb$ sample by about 18\% while retaining 98.5\% of the signal decays. 

In some cases, a single particle may be reconstructed as two nearly overlapping track segments in the vertex detector (VELO), which match two distinct track segments in the downstream tracking stations, producing two particles, one real and one fake. Such overlaps are highly unlikely in the decays studied here. Candidates having any pair of tracks with an opening angle in the VELO detector of less than 0.5~mrad are removed. The efficiency of this requirement on simulated $\Lb$ ($\Xibm$) decays is 99.8\% (99.7\%), while the background is suppressed by 4\% (30\%).

Signal $H_b$ candidates are partitioned into two subsamples based on the information from the L0 hardware trigger. The first category,
``triggered on signal'' (TOS), contains those candidates in which one or more of the final-state hadrons from the signal decay are associated with a positive L0 hadron trigger decision. The second category, ``triggered independently of the signal'' (TIS), represents the case where none of the final-state hadrons from the signal decay are associated with a positive L0 trigger decision. An event can satisfy both the TOS and TIS requirements, in which case the event is classified as TOS. The TOS signal decays tend to have larger $\pt$ than those of the TIS candidates due to the high energy transverse to the beamline that is required by the hadron trigger. Requiring events to be in one of these two categories retains 96\% of the $\Lb$ and $\Xibm$ signal decays selected by any L0 trigger in the data. At the level of the software trigger, the signal candidates are required to satisfy the requirements of the topological trigger~\cite{BBDT,LHCb-PROC-2015-018}.

\subsection{Simulation corrections}
\label{sec:simcor}
\begin{figure}[tb]
\centering
\includegraphics[width=0.96\textwidth]
{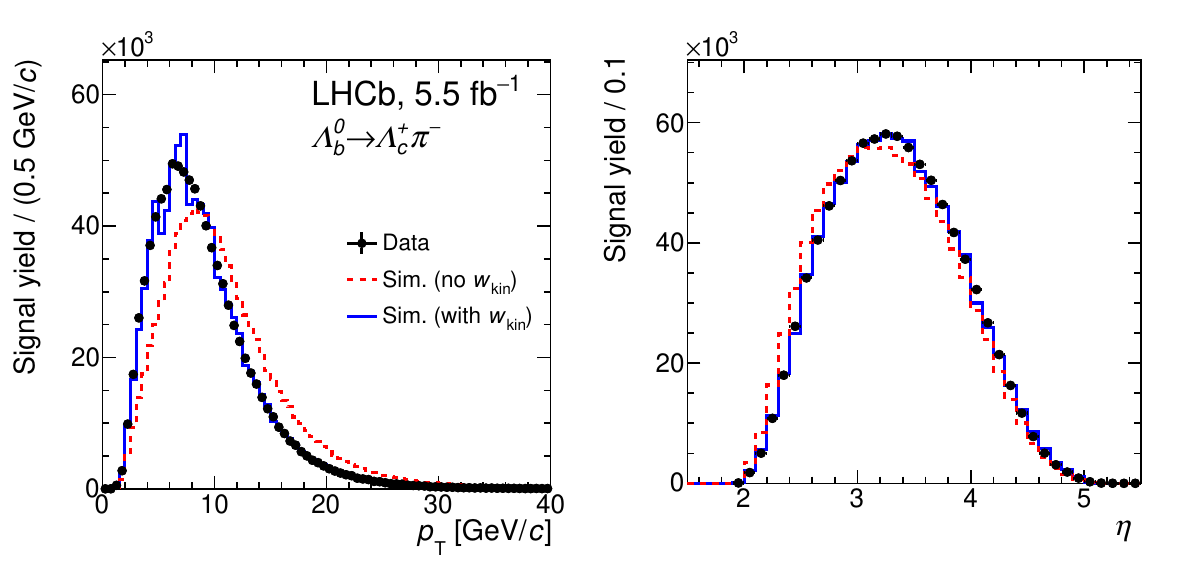}
\includegraphics[width=0.96\textwidth]{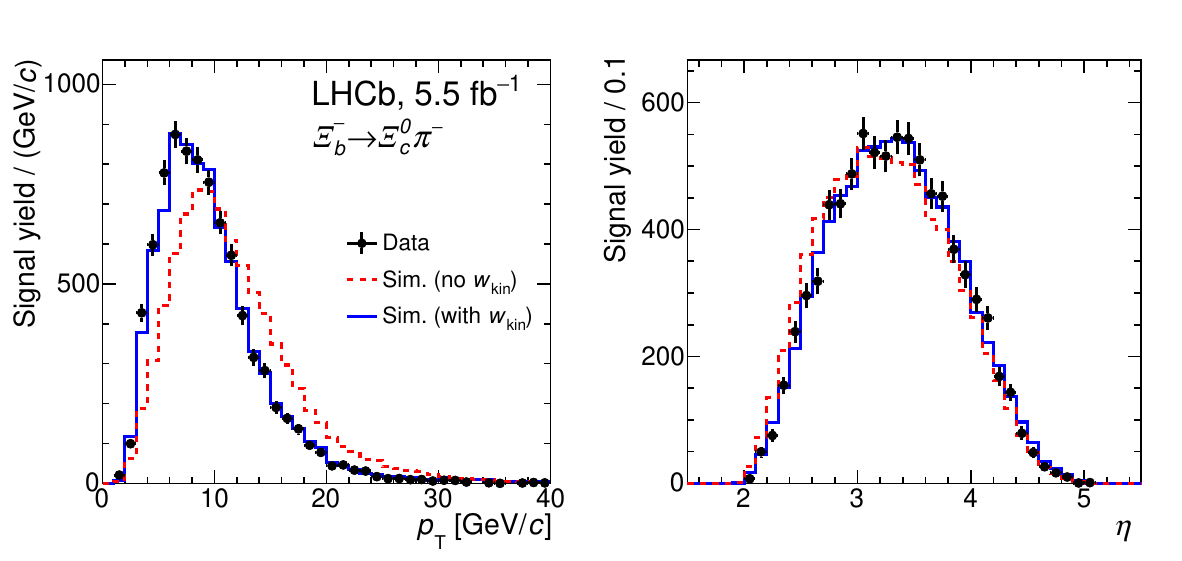}
\caption{\small{Comparison of the (left) $\pt$ and (right) $\eta$ spectra between background-subtracted data (points) and simulation (Sim.), before (red dashed line), and after (blue line) the kinematic weights ($w_{\rm kin}$) are applied. The top row shows the distributions for the $\Lb$ decay mode and the bottom shows those for the $\Xibm$ mode.
}}
\label{fig:HbKinWeights}
\end{figure}

A set of weights and corrections need to be applied to the simulation to account for imperfect modeling of the signal decays, the detector response and the relative luminosity for each year. The weights that are applied to the simulation correct for the $H_c$ decay kinematics, the $\Lb$ baryon lifetime, the $(\pt,\eta)$ spectra of the $H_b$ baryons and the fractions of TOS and TIS events. The weights are obtained using the selections described previously, before applying the multivariate discriminant, which is described later. Weights to account for differences in the $H_c$ decay kinematics 
are obtained from previous studies of semileptonic $\Lb\to\Lc\mun\neumb X$ and $\Xibm\to\Xicz\mun\neumb X$ decays
~\cite{LHCb-PAPER-2019-008}. The $\Lb$ lifetime weight accounts for a small difference between the value of $\tau_{\Lb}$ used in the simulation (1.451~ps) and the world average value of 1.464~ps. The $(\pt,\eta)$ weights ($w_{\rm kin}$)  for $H_b$ are obtained by taking the ratio of the background-subtracted signal distributions in the data~\cite{Pivk:2004ty} to the corresponding distributions in simulation, separately for each L0 trigger category. With those weights applied, the TOS fraction in simulation is about 20\% higher relative to that seen in the data. Weights are applied to the simulation so that the fractions of TOS and TIS events match those in the data. Projections of the $\pt$ and $\eta$ spectra in data and simulation with all  weights applied are shown in Fig.~\ref{fig:HbKinWeights}. Without the kinematic weights, the $\pt$ spectrum from the simulation is clearly shifted toward larger values than that of the data. After the weights are applied, the simulation matches the data well.

In addition to the weights, a correction is applied to the 2017 and 2018 simulations. With the increased radiation exposure in the VELO detector, cluster sizes increased due to more charge sharing between adjacent strips,  which in turn improved the IP resolution in data as compared to the simulation. Since quantities such as 
$\chisqip$ are used in the selection, and affect the decay-time acceptance, it is important that the simulation is well calibrated. Correction factors that depend on $p$ and $\eta$ are derived using the $\Lb\to\Lc\pim$ decay mode. This decay mode is ideal since 
the $\Lb$ baryon lifetime is precisely known,  and the signal yield is large ($\sim$$10^6$) with a high signal-to-background ratio. The derived corrections increase with particle momentum, and depend on $\eta$, due to the radiation profile~\cite{VELO_RadDamage}.
The correction to the $\chisqip$ value varies from 1.0, at low momentum, to about 1.55, on average, for protons with $p=200$\gevc. For the $\Km$, $\pip$ (from the $\Lc$ decay) and $\pim$ mesons (from the $\Lb$ decay) the average corrections at 200\gevc are about 1.35, 1.25 and 1.30, respectively.

\begin{figure}[tb]
\centering
\includegraphics[width=0.96\textwidth]{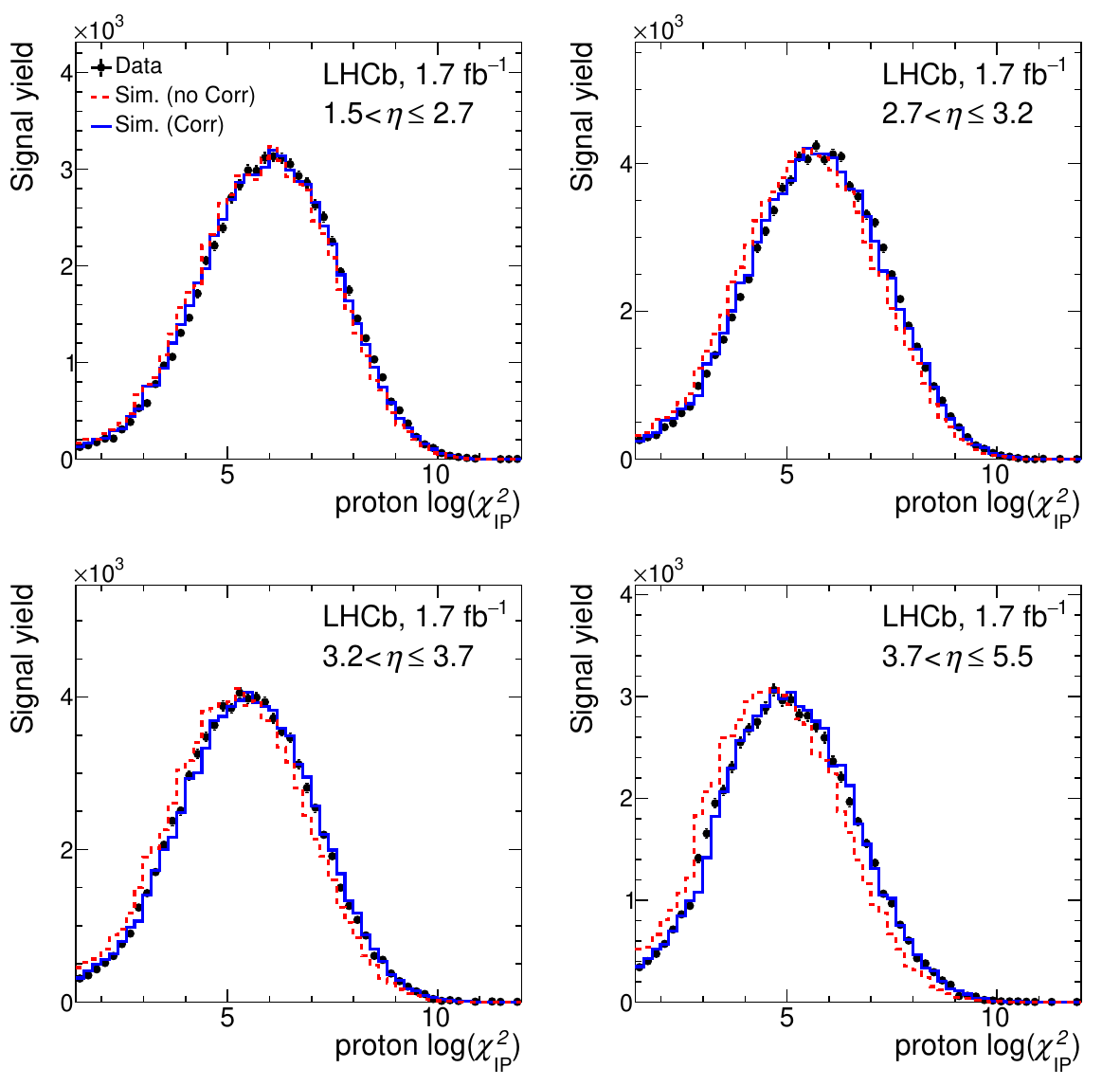}
\caption{\small{Distributions of $\log(\chisqip)$ for background-subtracted 2017 data (points), uncalibrated simulation [Sim. (no Corr)] (red dashed line), and calibrated simulation [Sim. (Corr)]  (blue solid line) for the proton from $\Lb\to\Lc\pim$ decays. The four plots show the distributions in four $\eta$ regions of the proton.
}}
\label{fig:p_2017_vs_eta}
\end{figure}

A comparison of the distributions of the $\log(\chisqip)$ of the proton from the $\Lc$ decay in four different $\eta$ regions is shown in Fig.~\ref{fig:p_2017_vs_eta} for the 2017 data and simulation. Prior to the correction, the simulation is clearly shifted with respect to the data, and the difference increases with $\eta$, as higher momentum tracks tend to be at smaller angles with respect to the beamline. After the correction, the simulation is in good agreement with the data. A similar level of agreement in $\chisqip$ is obtained for all final-state particles. 

Similar corrections are derived for the $H_b$ and $H_c$ vertex $\chisq$ values, which are unity at low momentum and increase to about 1.15 at 200\gevc. The calibrations are validated by comparing the background-subtracted distributions in data with the fully weighted simulation. The derived calibration parameters are then applied to the $\Xibm$ sample as well.

The hadron PID response of the final-state particles is calibrated using large samples of $\Dstarp\to(\Dz\to\Km\pip)\pip$ and ${\Lb\to\Lc\pim}$ decays as described in Ref.~\cite{LHCb-DP-2018-001}. Using the calibration data, the PID response for each hadron type is parameterized as a function of its transverse and total momentum, and the track multiplicity in the event~\cite{Poluektov_2015}. The PID response of each hadron in simulated signal decays is updated to use the response obtained from these calibration data, chosen at random from the relevant PID probability distribution.

\subsection{Multivariate selection}
To suppress background caused by random combinations of reconstructed particles, a gradient-boosted decision tree (BDT) algorithm~\cite{AdaBoost,Hocker:2007ht} is employed. The variables used to train the BDT classifiers include the $H_b$ decay vertex-fit $\chisq$ and $\cos(\theta_{\rm PV})$, where $\theta_{\rm PV}$ is the angle between the $H_b$ momentum vector and the vector that joins the PV and $H_b$ decay vertex; the $H_c$ decay vertex-fit $\chisq$ and the $H_c$ baryon decay time; and for each final-state particle, the $p$, $\pt$, $\chisqip$ values and a PID variable providing a measure of the likelihood that the PID information is consistent with the particle hypothesis~\cite{LHCb-DP-2018-001}.

A total of 20 (24) variables are used in the training of the $\Lb$ ($\Xibm$) BDT classifiers. The signal distributions are taken from simulated signal decays with all weights applied.  The background sample for the $\Lb$ ($\Xibm$) mode is taken from the high-mass sideband region, 5750--5950\mevcc (5900--6100\mevcc), after all selections are applied. For the $\Xibm$ decay, the $\Xicz$ mass requirement is relaxed to include candidates with ${|m(p\Km\Km\pip)-m_{\Xicz}|<30\mevcc}$ to increase the size of the background sample. A loose requirement is applied to the BDT algorithm output, which provides a signal efficiency of about 99\%, while suppressing the combinatorial background by a factor of four (six) for the $\Lb$ ($\Xibm$) decay mode. After the BDT requirement, less than 1\% of events have more than one $H_b$ candidate, and all candidates are retained.

\subsection{Fits to the mass distributions}
\begin{figure}[tb]
\centering
\includegraphics[width=0.48\textwidth]{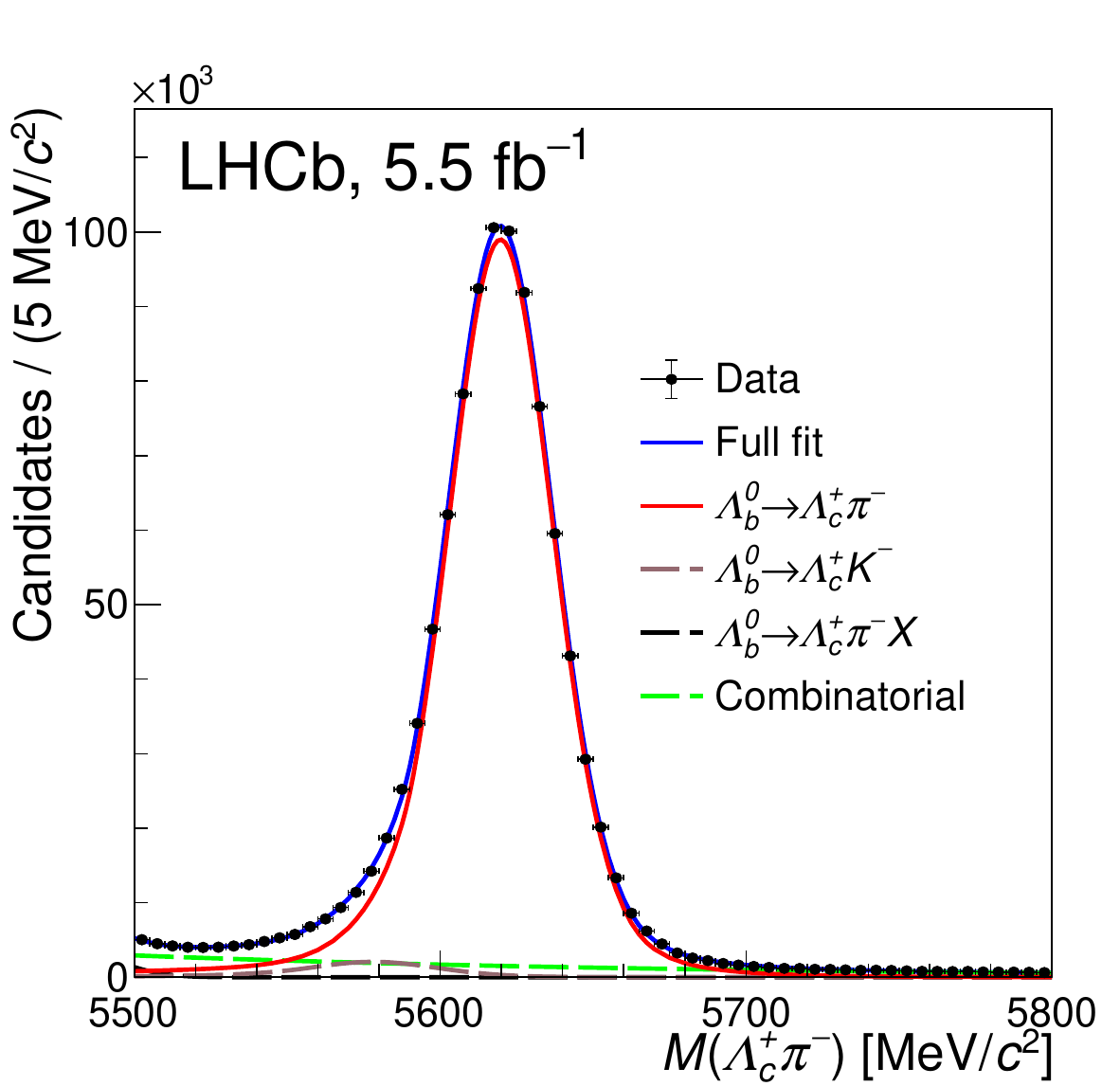}
\includegraphics[width=0.48\textwidth]{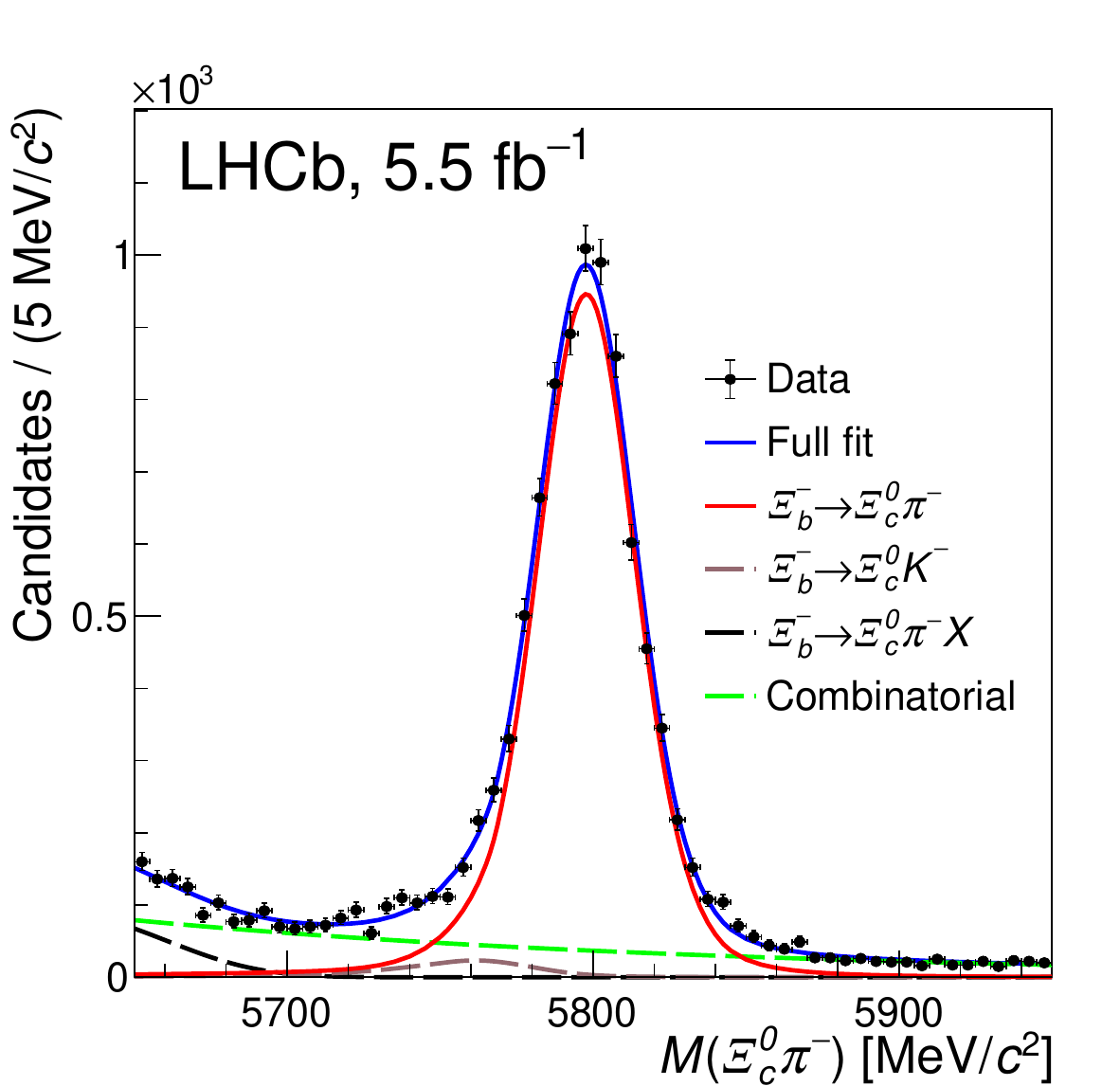}
\caption{\small{Invariant-mass spectra for (left) $\Lb\to\Lc\pim$ and (right) $\Xibm\to\Xicz\pim$ candidates in the data. The fit results, as described in the text, are overlaid.}} 
\label{fig:MassFitsFullSel}
\end{figure}

The invariant-mass spectra for selected candidates are shown in Fig.~\ref{fig:MassFitsFullSel}, along with the results of binned extended maximum-likelihood fits~\cite{BARLOW1990496}. Each mass spectrum is described by the sum of a signal function and three background shapes. The signal-mass shapes are parameterized by the sum of two Crystal Ball functions~\cite{Skwarnicki:1986xj} with a common mean. The signal shape parameters are fixed to the values obtained from simulation, except for the peak position and an overall scale factor, which accounts for a small difference in the mass resolution between data and simulation. In fits to the data, the scale factor is about 1.13, which is consistent with other analyses of $b$ hadron in LHCb.

The mass shape of misidentified $H_b\to H_c\Km$ decays is also described using the sum of two Crystal Ball functions, with shape parameters fixed to the values obtained from simulated $\Lb\to\Lc\Km$ decays. Taking into account the ratio of branching fractions 
$\BR(\Lb\to\Lc\Km)/\BR(\Lb\to\Lc\pim)$~\cite{PDG2022} and the relative selection efficiency, it is estimated that the yield fraction $N(H_b\to H_c\Km)/N(H_b\to H_c\pim)$ is $(3.1\pm0.6)\%$ for the $\Lb$ mode, and $(3.1\pm0.9)\%$ for the $\Xibm$ decay. An additional 20\% relative uncertainty is included for the $\Xibm$ mode to account for an assumption that the relative branching fraction between the Cabibbo-suppressed and Cabibbo-favored mode is equal to that of the $\Lb$ baryon~\cite{PDG2022}. These fractions are included in the fit with Gaussian constraints. Toward the lower edges of the mass distributions there are small contributions from partially reconstructed $b$-hadron decays with a missing pion or photon(s). The partially reconstructed background is modeled with an ARGUS shape~\cite{ARGUS:1990jet} convolved with a Gaussian resolution function. The shape parameters are obtained from a fit to the data in a wider mass region, and then fixed to those values in the default mass fits. 
The default mass fit includes candidates in the mass ranges 5500--5800\mevcc and 5650--5950\mevcc for the $\Lb$ and $\Xibm$ samples, respectively, which eliminates most of the partially reconstructed decays. The combinatorial background is described by an exponential function with the shape parameter free to vary in the fit. The results of the fit are overlaid on the data in Fig.~\ref{fig:MassFitsFullSel}. The $\Xibm$ and $\Lb$ signal yields for each L0 category and both L0 categories combined are given in Table~\ref{tab:HbYieldsWithBDT}.

\begin{table*}[ht]
\begin{center}
\caption{\small{Signal yields by L0 trigger category for $\Lb$ and $\Xibm$ decays for the total 2016--2018 data set after all selections. Each yield is obtained from an independent fit to the respective invariant-mass spectrum.}}
\begin{tabular}{lcccc}
\hline\hline
\\[-2.0ex]
Mode         &   TOS  &   TIS   & TOS + TIS \\
\hline          
\\[-2.2ex]
$\Xibm$     &      $4363\pm76$     &   $3976\pm71$  &   $~\,8303\pm107$ \\
\\[-2.2ex]
\hline
\\[-2.2ex]
$\Lb$ ($\times10^3$) &    $519.9\pm0.9$   &  $408.0\pm0.8$ &  $928.4\pm1.2$ \\
\\[-2.2ex]
\hline\hline
\end{tabular}
\label{tab:HbYieldsWithBDT}
\end{center}
\end{table*}

\section{Determination of $r_{\tau}$}
\begin{figure}[tb]
\centering
\includegraphics[width=0.49\textwidth]{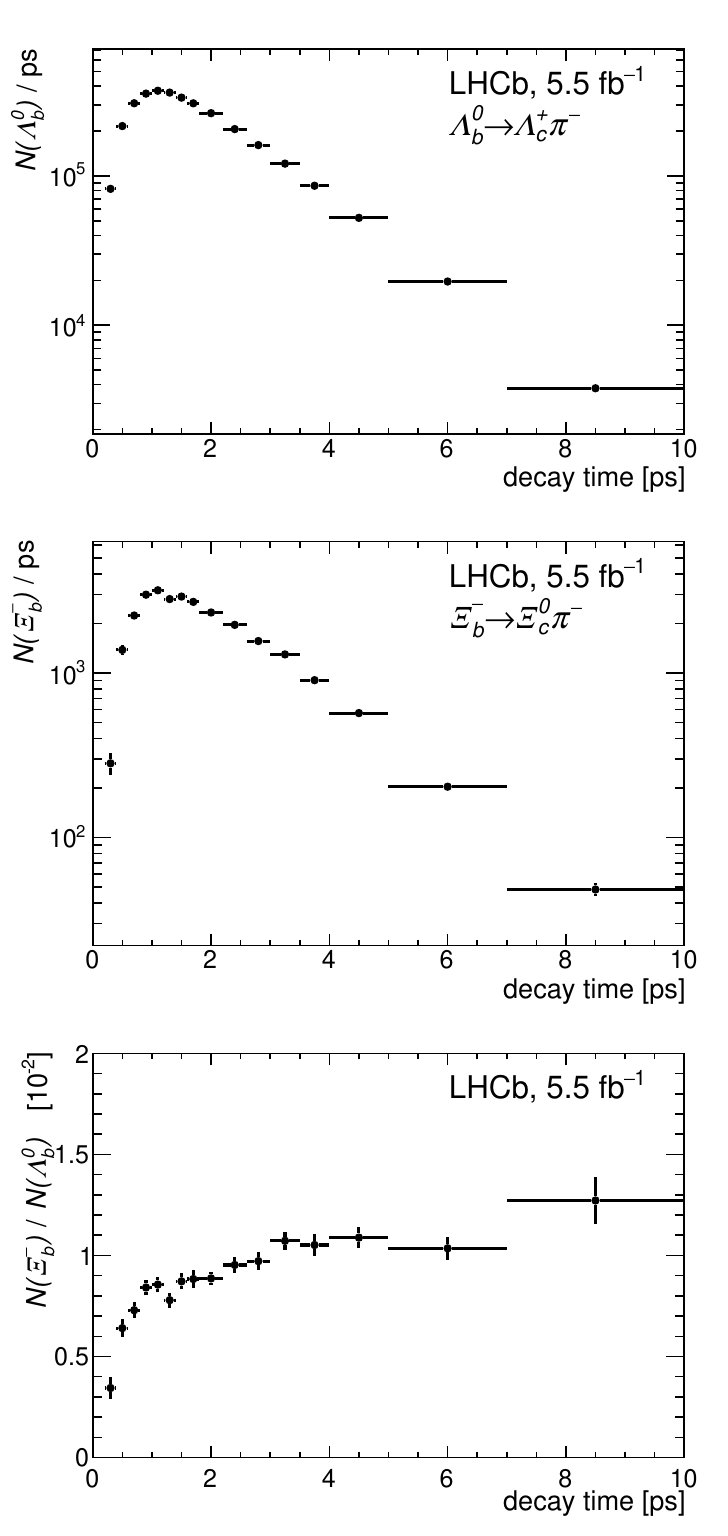}
\includegraphics[width=0.49\textwidth]{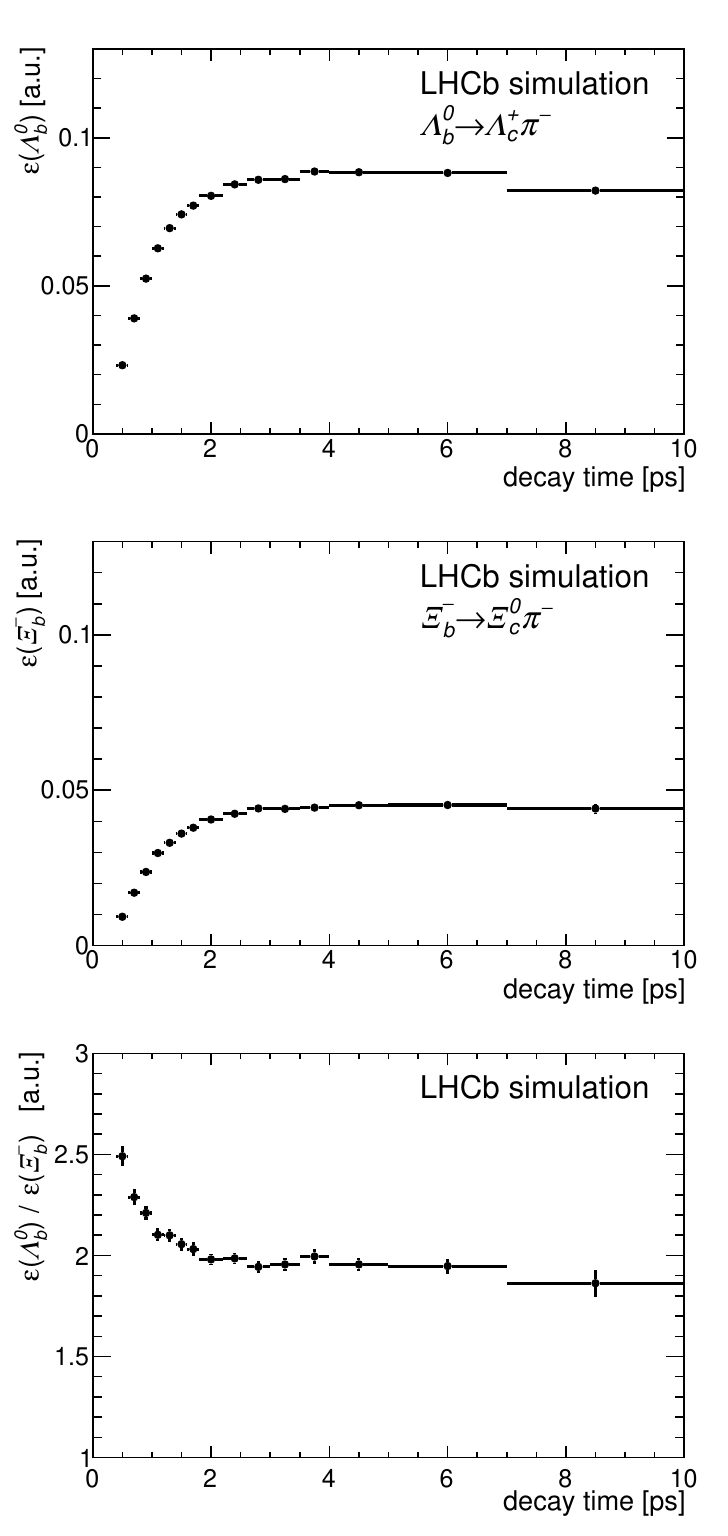}
\caption{\small{(Left) Signal yields divided by the bin width for the (top) $\Lb$ mode, (middle) $\Xibm$ mode as a function of decay time, and (bottom) the ratio $N(\Xibm)/N(\Lb)$. (Right) Selection efficiencies for the (top) $\Lb$ mode (middle) $\Xibm$ mode, and (bottom) the ratio of efficiencies $\eff(\Lb)/\eff(\Xibm)$, obtained from the weighted simulation. The uncertainties shown are due to the finite number of events in the data and simulation samples.}}
\label{fig:RawYieldsSum}
\end{figure}

The determination of the lifetime ratio $r_{\tau}$ requires the ratio of signal yields and the ratio of selection efficiencies, as shown in Eq.~\ref{eq:RT}.
The yields $N[\Xibm\to\Xicz\pim](t)$ and $N[\Lb\to\Lc\pim](t)$
are determined by fitting the combined TOS+TIS $H_b$ mass spectra in decay-time bins, ranging from 0.4 to $10\ps$. Specifically, the bin widths are $0.2\ps$ from 
$[0.4,\,1.8]\ps$, $0.4\ps$ from $[1.8,\,3.0]\ps$, $0.5\ps$ from $[3,\,4]\ps$,
and 1 bin each from [4,\,5], [5,\,7] and $[7,\,10]\ps$. The mass fits are performed as described previously, where the signal shape parameters in each decay-time bin are obtained from the corresponding fit to the simulated signal decays in the same decay-time bin, with all weights applied. The mass resolution scale factor is Gaussian constrained to the value obtained from the fit to the full data sample. This is validated using simulation, where the mass resolution is found to be independent of the decay time. All signal and background yields are free parameters in the fit for each decay-time bin. The yields of $\Lb$ and $\Xibm$ baryon decays as a function of the decay time are shown in Fig.~\ref{fig:RawYieldsSum}\,(left), along with the ratio of yields, $N(\Xibm)/N(\Lb)$. The yield ratio decreases as the decay time approaches zero. This trend is expected, and is due to the detachment requirement (see Section~\ref{sec:selection}) on all final-state tracks from all PVs, which has a lower efficiency for the five-particle final state ($\Xibm$) than the four-particle final state ($\Lb$).

The selection efficiencies for decays within the LHCb acceptance are obtained from simulation, corrected for data-simulation discrepancies as detailed in Section~\ref{sec:simcor}. The resulting efficiencies versus decay time are shown in Fig.~\ref{fig:RawYieldsSum}\,(right). The $\Lb$ and $\Xibm$ signal efficiencies show a steep rise from zero decay time, followed by a plateau. Due to very low selection efficiency, and potentially large systematic effects, only $H_b$ decay times larger than 0.4~ps are used in the lifetime determination. As shown in the bottom plots of Fig.~\ref{fig:RawYieldsSum}, the reduction in the yield ratio $N(\Xibm)/N(\Lb)$ mirrors the ratio of selection efficiencies $\eff(\Lb)/\eff(\Xibm)$.

\begin{figure}[tb]
\centering
\includegraphics[width=0.80\textwidth]{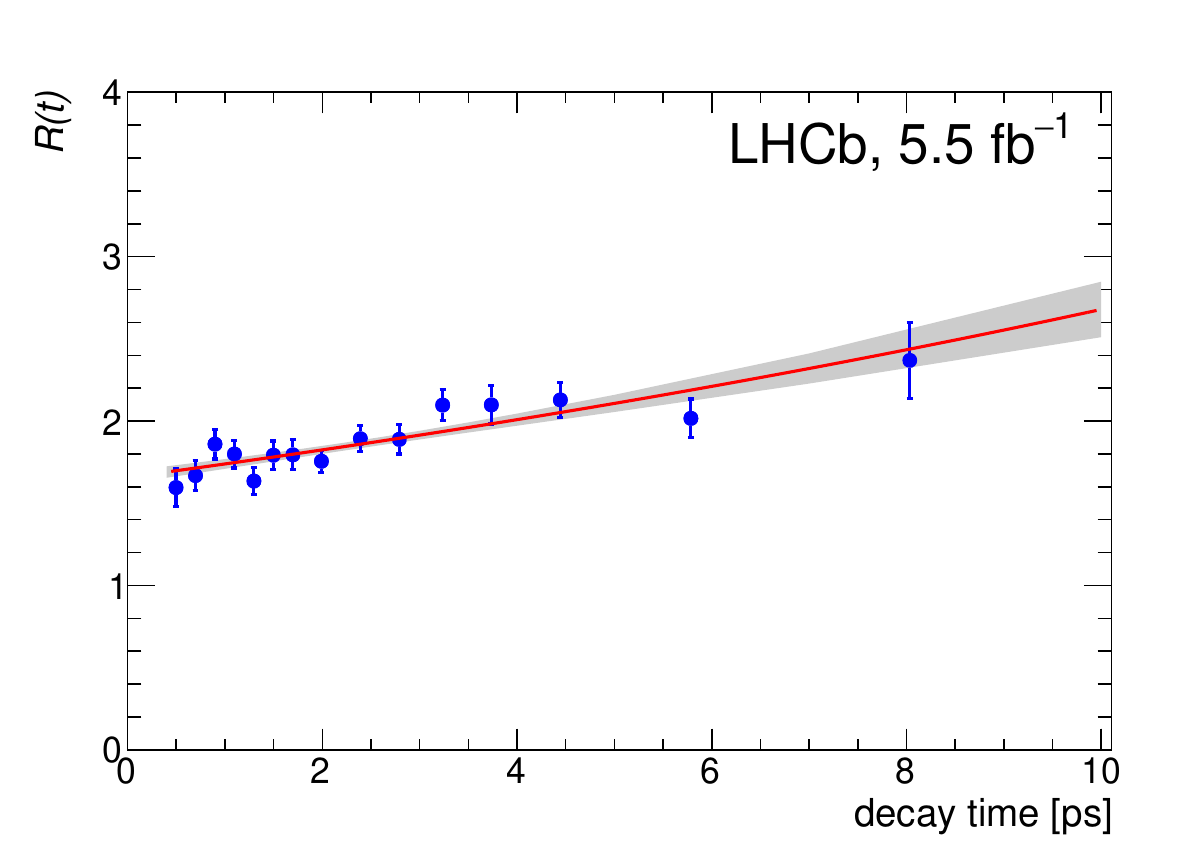}
\includegraphics[width=0.80\textwidth]{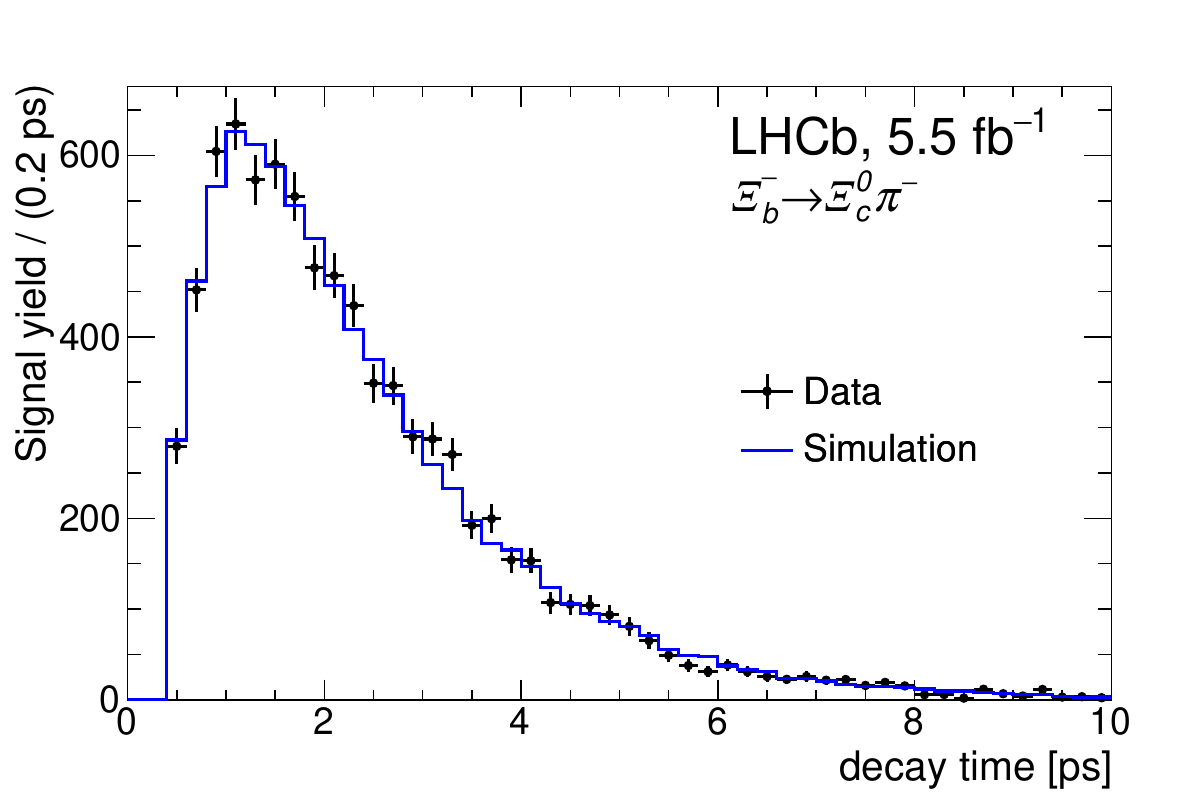}
\caption{\small{(Top) Corrected yield ratio $R(t)$ as a function of the decay time. The red line shows the fit result, and the gray band shows the 68\% confidence level interval. (Bottom) Overlay of the decay-time spectrum of $\Xibm$ signal decays and simulation using the best fit lifetime of 1.575~ps.
}}
\label{fig:CorrYieldRatio}
\end{figure}

The product of the yield and efficiency ratios shown in Fig.~\ref{fig:RawYieldsSum}\,(bottom) gives $R(t)$, and the result is shown in Fig.~\ref{fig:CorrYieldRatio}\,(top). The points are placed along the time axis at the weighted average within the bin, assuming a decay-time spectrum that is exponential with an effective lifetime of 1.51~ps, which is halfway between the world-average values of the $\Lb$ and $\Xibm$ baryons' lifetimes. The results are insensitive to choosing other values within the range from 1.47 to 1.61~ps. The results of a fit to an exponential function are overlaid on the data, yielding $\lambda=(48.3\pm7.8)\times10^{-3}~\ps^{-1}$, where the uncertainty is due to the finite signal yields in the data. From this value, the relative lifetime and $\Xibm$ lifetime are readily computed as
\begin{align}
    r_{\tau}^{\rm Run\,2}&=1.076\pm0.013, \nonumber\\
    \tau_{\Xibm}^{\rm Run\,2}&=1.575\pm0.019~{\rm ps},\nonumber
\end{align}
\noindent where the uncertainties are statistical only.

Figure~\ref{fig:CorrYieldRatio}\,(bottom) shows an overlay of the decay-time spectrum of $\Xibm$ data obtained using the \sPlot method~\cite{Pivk:2004ty}, and simulation weighted to represent the best fit $\Xibm$ lifetime of 1.575~ps. It is seen that the best fit lifetime is in good agreement with the data.

\section{Systematic uncertainties and cross-checks}
A number of possible sources of systematic uncertainty are considered and shown in Table~\ref{tab:Syst}, along with their associated values. For each of the sources, unless otherwise indicated, the systematic uncertainty is assessed by making a change to the default method, and assigning the relative change in $r_{\tau}$ with respect to the default value as the systematic uncertainty. The total is obtained from the quadrature sum of the values from each source.

\begin{table*}[tb]
\begin{center}
\caption{\small{Sources and values of the relative systematic uncertainty on $r_{\tau}$.}}
\begin{tabular}{lc}
\hline\hline
Source                    &   Value (\%) \\
\hline
Simulated sample size     &     0.43     \\
Signal shape              &     0.07     \\           
Background shape          &     0.01     \\  
$\chisqip$ scaling        &     0.20    \\
Truth matching            &     0.07     \\    
Bin width in mass         &     0.03     \\  
Mass fit range             &     0.18 \\    
Bin width in time         &     0.06     \\   
BDT requirement           &    0.21     \\  
$\Lb$ lifetime            &     0.05 \\
\hline
Total                     &     0.57 \\
\hline\hline
\end{tabular}
\label{tab:Syst}
\end{center}
\end{table*}

The largest systematic uncertainty is due to the finite simulated sample sizes. The uncertainty is obtained by performing 1000 alternative fits for $r_{\tau}$, where in each fit the relative efficiency in each bin is fluctuated by its uncertainty according to a Gaussian distribution. The standard deviation of the 1000 alternative fits is assigned as the systematic uncertainty.

The uncertainty due to the assumed signal shape is quantified by using the sum of a Bukin~\cite{Verkerke:2003ir} and Gaussian function with a common peak value for the signal function to fit the data. Sensitivity to the choice of the combinatorial background function is investigated by using a third-order Chebychev polynomial instead of an exponential function in the mass fits. 

To assess the sensitivity to the corrections applied to the $\chisqip$ observable of final-state tracks in the 2017 and 2018 simulations, the parameters used for the scaling are modified from their best values by about one standard deviation. The selections are reapplied, and the full analysis is rerun with the new scaling parameters.

For the efficiency determination, the reconstructed signal decays in the simulation are required to be truth-matched with the generated signal decays. In a few percent of cases, this association can fail if the fraction of matched hits between the reconstructed track and the true particle do not exceed a minimum threshold. The potential impact is assessed by removing the matching requirement entirely for the efficiency determination.

The mass fits are binned extended maximum-likelihood fits, with bin widths of 2.0 and 5.0\mevcc for the $\Lb$ and $\Xibm$ mass fits, respectively. To assess whether the choice of binning has any impact on the results, the bin widths in mass are halved,
and new values of $N[\Xibm\to\Xicz\pim](t)$ and $N[\Lb\to\Lc\pim](t)$ are used in place of the default ones. To study the sensitivity to the range in mass over which the mass spectrum is fitted, the region of the mass fit is extended by 50\mevcc on both sides and the fits are redone.

To study the effects of the binning in decay time, the number of decay-time bins is increased from 16 to 25. The mass fits in the narrower decay-time bins are redone, leading to a new set of data points for $R(t)$, and a corresponding change in the fitted $r_{\tau}$. 

The BDT requirement is highly efficient, about 99\% on simulated signal decays. Nonetheless, some of the inputs are correlated with decay time. To estimate a potential impact of the BDT requirement on the measurement of $r_{\tau}$, the operating point is made less stringent, doubling the amount of combinatorial background, and increasing the signal efficiency to at least 99.5\%. The observed change is assigned as a systematic uncertainty, although it certainly includes a statistical component due to the increased level of background in the mass fits.

The systematic uncertainty in $r_{\tau}$ due to the limited precision of the measured $\Lb$ baryon lifetime is
small, due to its precisely known value. The total systematic uncertainty on $r_{\tau}$ is 0.57\%, which can be compared to the statistical precision of about 1.2\%. 

A number of cross-checks are performed to ensure self-consistency among different subsets of the data. The data and simulation are divided into mutually exclusive data sets, and the lifetime is measured separately in each subsample. Four different partitions are investigated: by year (2016, 2017, and 2018), by L0 trigger category (TOS and TIS), by LHCb magnet polarity, and by track multiplicity in the $pp$ collision event ($N_{\rm trk}\leq150$ and $N_{\rm trk}>150$). All of the lifetime values obtained from the twelve independent subsets are within one standard deviation of the overall average.

\section{Summary}
In summary, using a $pp$ collision data sample corresponding to an integrated luminosity of 5.5\invfb, the LHCb collaboration measures the ratio of lifetimes of the $\Xibm$ baryon relative to that of the $\Lb$ baryon to be
\begin{align*}
    r_{\tau}^{\rm Run\,2} = 1.076\pm0.013\pm0.006,
\end{align*}
\noindent where the uncertainties are statistical and systematic, respectively. Multiplying by the known $\Lb$ baryon lifetime of $1.464\pm0.010$~ps~\cite{PDG2022}, the $\Xibm$ lifetime is found to be
\begin{align*}
    \tau_{\Xibm}^{\rm Run\,2} = 1.575\pm0.019\pm0.009\pm0.011~{\rm ps},
\end{align*}
\noindent where the last uncertainty is due to that of the $\Lb$ baryon lifetime. This is the most precise measurement of the $\Xibm$ baryon lifetime and is consistent with previous measurements.

The $r_{\tau}$ value obtained here is combined with the corresponding value obtained with the Run\,1 data sample of $1.089\pm0.026\pm0.011$~\cite{LHCb-PAPER-2014-048}. The resulting Run\,1 and Run\,2 average values are 
\begin{align*}
r_{\tau}^{\rm Run\,1,2} &= 1.078\pm0.012\pm0.007,\\
\tau_{\Xibm}^{\rm Run\,1,2} &= 1.578\pm0.018\pm0.010\pm0.011~{\rm ps}.
\end{align*}
All uncertainties, except that of the $\Lb$  baryon lifetime, are taken as uncorrelated in the average.
This represents a statistical precision on the lifetime ratio of about 1.3\%, which improves on the world-average value by about a factor of two. The measured value is in agreement with the prediction from the heavy quark expansion, of $r_{\tau}^{\rm HQE}=1.078\pm0.021$~\cite{Gratrex}. This prediction only accounts for the decay width of the $b$ quark and higher-order corrections. Decays of the $\Xibm$ baryon involving the weak $s\to u\bar{u}d$ transition have been observed by LHCb~\cite{LHCb-PAPER-2023-015}, with branching fraction $\BR(\Xibm\to\Lb\pim)\sim1\%$, which would reduce the central value of the theoretical prediction by about 1\%. After applying this correction the theoretical calculation would still be in agreement with the measured $r_{\tau}$ value.
\clearpage
\section*{Acknowledgements}
%
%
\noindent We express our gratitude to our colleagues in the CERN
accelerator departments for the excellent performance of the LHC. We
thank the technical and administrative staff at the LHCb
institutes.
We acknowledge support from CERN and from the national agencies:
CAPES, CNPq, FAPERJ and FINEP (Brazil); 
MOST and NSFC (China); 
CNRS/IN2P3 (France); 
BMBF, DFG and MPG (Germany); 
INFN (Italy); 
NWO (Netherlands); 
MNiSW and NCN (Poland); 
MCID/IFA (Romania); 
MICINN (Spain); 
SNSF and SER (Switzerland); 
NASU (Ukraine); 
STFC (United Kingdom); 
DOE NP and NSF (USA).
We acknowledge the computing resources that are provided by CERN, IN2P3
(France), KIT and DESY (Germany), INFN (Italy), SURF (Netherlands),
PIC (Spain), GridPP (United Kingdom), 
CSCS (Switzerland), IFIN-HH (Romania), CBPF (Brazil),
and Polish WLCG (Poland).
We are indebted to the communities behind the multiple open-source
software packages on which we depend.
Individual groups or members have received support from
ARC and ARDC (Australia);
Key Research Program of Frontier Sciences of CAS, CAS PIFI, CAS CCEPP, 
Fundamental Research Funds for the Central Universities, 
and Sci. \& Tech. Program of Guangzhou (China);
Minciencias (Colombia);
EPLANET, Marie Sk\l{}odowska-Curie Actions, ERC and NextGenerationEU (European Union);
A*MIDEX, ANR, IPhU and Labex P2IO, and R\'{e}gion Auvergne-Rh\^{o}ne-Alpes (France);
AvH Foundation (Germany);
ICSC (Italy); 
GVA, XuntaGal, GENCAT, Inditex, InTalent and Prog.~Atracci\'on Talento, CM (Spain);
SRC (Sweden);
the Leverhulme Trust, the Royal Society
 and UKRI (United Kingdom).




\clearpage
\addcontentsline{toc}{section}{References}
\bibliographystyle{LHCb}
\bibliography{main,standard,LHCb-PAPER,LHCb-CONF,LHCb-DP,LHCb-TDR}

\newpage
\centerline
{\large\bf LHCb collaboration}
\begin
{flushleft}
\small
R.~Aaij$^{36}$\lhcborcid{0000-0003-0533-1952},
A.S.W.~Abdelmotteleb$^{55}$\lhcborcid{0000-0001-7905-0542},
C.~Abellan~Beteta$^{49}$,
F.~Abudin{\'e}n$^{55}$\lhcborcid{0000-0002-6737-3528},
T.~Ackernley$^{59}$\lhcborcid{0000-0002-5951-3498},
A. A. ~Adefisoye$^{67}$\lhcborcid{0000-0003-2448-1550},
B.~Adeva$^{45}$\lhcborcid{0000-0001-9756-3712},
M.~Adinolfi$^{53}$\lhcborcid{0000-0002-1326-1264},
P.~Adlarson$^{79}$\lhcborcid{0000-0001-6280-3851},
C.~Agapopoulou$^{13}$\lhcborcid{0000-0002-2368-0147}
% ...

{\footnotesize \it

% ...
$^{13}$Universit{\'e} Paris-Saclay, CNRS/IN2P3, IJCLab, Orsay, France\\
% ...
$^{36}$Nikhef National Institute for Subatomic Physics, Amsterdam, Netherlands\\
% ...
$^{45}$Instituto Galego de F{\'\i}sica de Altas Enerx{\'\i}as (IGFAE), Universidade de Santiago de Compostela, Santiago de Compostela, Spain\\
% ...
$^{49}$Physik-Institut, Universit{\"a}t Z{\"u}rich, Z{\"u}rich, Switzerland\\
% ...
$^{53}$H.H. Wills Physics Laboratory, University of Bristol, Bristol, United Kingdom\\
% ...
$^{55}$Department of Physics, University of Warwick, Coventry, United Kingdom\\
% ...
$^{58}$School of Physics and Astronomy, University of Glasgow, Glasgow, United Kingdom\\
$^{59}$Oliver Lodge Laboratory, University of Liverpool, Liverpool, United Kingdom\\
% ...
$^{67}$Syracuse University, Syracuse, NY, United States\\
% ...
$^{79}$Department of Physics and Astronomy, Uppsala University, Uppsala, Sweden, associated to $^{58}$\\
% ...
}
\end{flushleft}

\end{document}